\documentclass[10pt]{article}
\usepackage[utf8]{inputenc}
\usepackage[english]{babel}
\usepackage{amsmath}
\usepackage{amsfonts}
\usepackage{bm}
\usepackage{amssymb}
\usepackage{amscd}
\usepackage{amsthm}

\usepackage{url}
\usepackage{graphicx}
\usepackage{subfig}
\usepackage{epsfig}
\usepackage{wrapfig}
\usepackage{pifont}
\usepackage{rotating}
\usepackage{titlesec}
\usepackage{fancyhdr}

\usepackage{xr}

\begin{document}

\markboth{ S. Cl\'emen\c{c}on \textit{et al.}}{\it A Network Analysis of the HIV/AIDS Epidemics in Cuba}

\title{A Statistical Network Analysis \\
of the HIV/AIDS Epidemics in Cuba}

\author{ St\'ephan Cl\'emen\c{c}on$^{1}$\footnote{Author for correspondence (stephan.clemencon@telecom-paristech.fr)} \and Hector De Arazoza$^{2, 3, 5}$ \and Fabrice Rossi$^{1,4}$ \and Viet Chi Tran$^{5}$\footnote{Author for correspondence (chi.tran@math.univ-lille1.fr)}}

\maketitle

\noindent \small{$^1$ Institut Telecom, LTCI UMR Telecom ParisTech/CNRS 5141, 46 rue Barrault, 75 634 Paris, France\\
$^2$ Universit\'e Ren\'e Descartes, MAP5 UMR CNRS 8145, 45 rue des Saints P\`eres, 75 270 Paris, France\\
$^3$ Facultad de Matem\'atica y Computaci\'on, Universidad de la Habana, La Habana, Cuba\\
$^4$ SAMM EA 4543, Universit\'e Paris 1 Panthéon-Sorbonne, Centre PMF, 90 rue de Tolbiac, 75013 PARIS Cedex 13, France\\
$^5$ Laboratoire Paul Painlev\'e, UMR CNRS No. 8524, Universit\'e Lille 1,  59 655 Villeneuve d'Ascq Cedex, France.
}


\begin{abstract}
The Cuban contact-tracing detection system set up in 1986 allowed the reconstruction and analysis of the sexual network underlying the epidemic (5,389 vertices and 4,073 edges, giant component of 2,386 nodes and 3,168 edges), shedding light onto the spread of HIV and the role of contact-tracing. Clustering based on modularity optimization provides a better visualization and understanding of the network, in combination with the study of covariates. The graph has a globally low but heterogeneous density, with clusters of high intraconnectivity but low interconnectivity. Though descriptive, our results pave the way for incorporating structure when studying stochastic SIR epidemics spreading on social networks.\end{abstract}

\begin{center}
{\bf Data accessibility: } Data concerning the sex, sexual orientation, mode of detection of individuals and allowing the reconstruction of the graph, is provided in the supplementary materials.\\
{\bf Keywords:}
Cuban HIV/AIDS epidemics; contact-tracing; social network; graph-mining; clustering.\\
\end{center}

\section{Introduction}

Since 1986, a contact-tracing detection system has been set up in Cuba in order to bring the spread of the HIV epidemics under control. It has also enabled the gathering of a considerable amount of detailed epidemiological data at the individual level. In the resulting database, any individual tested as HIV positive is indexed and, for confidentiality reasons, described through several attribute variables: gender, sexual orientation, way of detection, age at detection, date of detection, area of residence, etc. The list of indices corresponding to the sexual partners appearing in the database she/he possibly named for contact-tracing is also available. It is worth recalling that in Cuba HIV spreads essentially through sexual transmission. Female to female transmission and infection by blood transfusion or related to drug use are neglected (e.g. \cite{chanetal}).  We refer to \cite{Auvert07} for a preliminary overview of the HIV/AIDS epidemics in Cuba, as well as a description and the context of the construction of the database used in the present study and the context in which it was constructed.\\
The main purpose of this paper is to reconstruct a graph of sexual partners that have been diagnosed HIV positive on the Cuban data repository and carry out an exploratory statistical analysis of the resulting sexual contact network. The network is composed of 5,389 vertices, or nodes, that correspond to the individuals diagnosed as HIV positive between 1986 and 2006 in Cuba, i.e. 1,109 women (20.58\%) and 4,280 men (79.42\%); 566 (10.50\%) of which are heterosexual and 3,714 (68.92\%) are Men who have Sex with Men (MSM in abbreviated form; men who reported at least one sexual contact with another man in the two years preceding HIV detection). Individuals declared as sexual contacts but who are not HIV positive are not listed in the database. The vertices that depict the fact that two individuals have been sexual partners during the two years that preceded the detection of either one are linked by 4,073 edges. Our data exhibit a ``giant component", counting 2,386 nodes. It is remarkable that in the existing literature on sexually transmitted diseases graph networks are generally smaller and/or do not exhibit such a large connected component and/or contain a very small number of infected persons (e.g. \cite{rothenbergwoodhousepotteratmuthdarrowklovdahl,wyliejolly}). \\

Recently, network concepts have received much attention in epidemiology, essentially for modeling purposes. References are far too numerous to be exhaustively listed here and we refer to e.g. Durrett \cite{durrett}, Newman \cite{Newman03}, House \cite{house} for comprehensive reviews on the topic. Mathematical models of epidemiological networks are obtained by  mean-field approximation (e.g. \cite{durrett,KG99,pastorsatorrasvespignani}) or through large population approximations (e.g. \cite{ballneal,decreusefonddhersinmoyaltran,grahamhouse,barbourreinert}) and generally stipulate simple structures for the network: small worlds (e.g. \cite{KG99, MN00}), configuration models (e.g. \cite{ML01, volz, volzancelmeyers,kissgreenkao}) and random intersection graphs (e.g. \cite{brittondeijfenlageraslindholm,ballsirltrapman}). The present study lies upstream of explicit network-based epidemic modeling and aims at quantitatively describing, the properties of the observed sexual contact network and explaining its complex structure. Recent graph-mining techniques are used in order to describe the connectivity/communication properties of the sexual contact network and understand the impact of heterogeneity (with respect to the attributes observed) in the graph structure. Particular attention is paid to the graphical representation of the data, as conventional methods cannot be used with databases of the size of the one used in this study. The idea is to obtain a clustering of the population so as to represent structural information in an interpretable way.
Beyond global graph visualization, the task of partitioning the network into groups, with dense internal links and low external connectivity, is known as clustering. In contrast to standard multivariate analysis, in which the network structure of the data is ignored, our method has shed light on how different mechanisms  (e.g. social behavior, detection system) have affected the epidemics of HIV in the past, and provide a way of predicting the future evolution of this disease. This study paves the way for building more realistic network models in the field of mathematical epidemiology. In the next few years, improvements in individual data collection will certainly give rise to an increasing number of empirical analyses based on these mathematical notions.

\section{Methods} 


Temporarily ignoring  the network structure, we first compute a collection of basic descriptive statistics so as to explore the dependence structures at the population level and provide a first overview of the underlying epidemics. This analysis is shown in the Electronic Supplementary Materials (ESM, \cite{clemenconarazozarossitranESM}).
The network structure is then taken into account and we compute a wide variety of statistics related to the degree distribution, geodesic distances, mixing and clustering patterns. These statistics are useful for summarizing the properties of a network that is too large to be visualized directly (see e.g. \cite{Newman03}).
Finally, graph-mining techniques are detailed and the statistical methods for visualizing a graph with a high number of vertices are carefully considered. Here, with a giant component of 2,386 nodes (44.28\% of total number of vertices) and 3,168 edges (77.78\% of the total number of edges), basic graph representation methods fail, showing very dense parts surrounded by "tree-like" subgraphs. To improve the graphical representations, optimization algorithms can be used to produce an aesthetically pleasing and possibly informative layout (with a compact representation involving a minimal number of crossing edges, a maximal display of symmetries and an even distribution of the vertices).
 Once a certain strength measure of connectivity has been given, it is generally turned into a NP-hard optimization problem, of which acceptably good solutions can be computed in a reasonable amount of time by means of metaheuristics such as simulated annealing, tabu search or genetic algorithms. In order to get an insight into the general organization of the largest connected component of the network, we divide it into several communities, so that nodes that are highly connected are grouped and those that are poorly connected are separated. The groups/communities thus defined are then characterised by a list of descriptive statistics. Homogeneity tests are performed in order to assess possible significant differences between these statistical subpopulations. The individuals corresponding to extra-group links are isolated.

\subsection{Degree distribution}\label{subsec:degree_dist}


We calculate the degree distribution $(p_k:\; k\in \mathbb{N})$ using the number of declared sexual partners in the two years preceding detection, where $p_k$ is the proportion of vertices having declared $k$ sexual partners. Other degree distributions (the observed number of neighbours in the graph which includes only HIV positive contacts or the number of tested partners) are studied in the ESM. The analysis of the degree distribution is carried out for specific types of edges, depending on the 'gender/sexual orientation' or the 'area of residence' or the 'age at detection', in order to shed light on possible heterogeneity in the population.

The degree distributions of most real-world networks, referred to as scale-free networks, often exhibit a power-law behavior in their right tails (see \cite{durrett}), meaning that
 \begin{equation*}\label{power_law}
 p_k\sim k^{-\alpha}, \text{ as } k \text{ becomes large},
 \end{equation*}
 for some exponent $\alpha> 1$ (notice that $\sum_{k=1}^{\infty}1/k^{\alpha}<\infty$ in this case). Roughly speaking, this describes the situations where the majority of vertices have few connections, but a small fraction of the vertices are highly connected (e.g. Chapter 4 in \cite{NBW06} for further details). We propose to fit a power-law exponent and consider two methods for this purpose, see also \cite{CSN09}. First, we  minimize, over $\alpha>1$, the following measure of dissimilarity between the observed degree distribution and the power-law distribution with exponent $\alpha$ based on the values larger than $k_0$ for the degree
 \begin{equation}\label{dissimilarity}
\mathcal{K}_{k_0}(p, \alpha)=\sum_{k\geq k_0}\frac{p_k}{c_{p,k_0}}\log\left(\frac{C_{\alpha}\cdot p_k}{c_{p,k_0}\cdot k^{-\alpha}}\right),
\end{equation}
where $\log$ denotes the natural logarithm, $c_{p,k_0}=\sum_{k\geq k_0}p_k$ and $C_{\alpha}=\sum_{k\geq k_0}1/k^{\alpha}$.
Notice that, when $k_0$ is larger than the maximum observed degree distribution $k_{\max}$, we have $\mathcal{K}_{k_0}(p, \alpha)=0$ no matter the exponent $\alpha$. Also, the computation of \eqref{dissimilarity} involves summing a finite number of terms only, since the empirical frequency $p_k$ is equal to zero for any degree $k$ sufficiently large. The criterion $\mathcal{K}_{k_0}(p, \alpha)$ is known as the Kullback-Leibler divergence between the empirical and theoretical conditional distributions given that the degree is larger than $k_0$. Incidentally, we point out that other dissimilarity measures could be considered for the purpose of fitting a power-law, such as the $\chi^2$-distance for instance. For a fixed threshold $k_0\geq 1$, it is natural to select the value of the power-law exponent that provides the best fit, that is:
\begin{equation*}\label{min_contrast}
\widehat{\alpha}_{k_0}=\arg\min_{\alpha>1}\mathcal{K}_{k_0}(p, \alpha).
\end{equation*}
Choosing $k_0$ precisely being a challenging question to statisticians. Following in the footsteps of the heuristic selection procedures proposed in the context of heavy-tailed continuous distributions (see Chapter 4 in \cite{Resnick}), when possible, we suggest to choose $\widehat{\alpha}_{k_0}$ with $k_0$ in a region where the graph $\{(k,\; \widehat{\alpha}_{k}):\; k=1,\;\ldots, \; k_{\max}\}$ is becoming horizontal, or at least shows an inflexion point. For completeness, we also compute the Hill estimator:
\begin{equation*}
\widetilde{\alpha}_{m}=\left( \frac{1}{m}\sum_{j=1}^{m}\frac{k_{(j)}}{k_{(m)}} \right)^{-1},\label{hill}
\end{equation*}
where $n$ is the number of vertices of the graph under study, $1\leq m\leq n$ and $k_{(1)}=k_{\max},\; k_{(2)},\; \ldots,\; k_{(m)}$ denote the $m$ largest observed degrees sorted in decreasing order of their magnitude. The tuning parameter $m$ is selected graphically, by plotting the graph $\{(m,\widetilde{\alpha}_m):\; m=1,\; \ldots,\; n\}$. In the case when the degrees of the vertices of the graph are independent, as for the configuration model \cite{NSW01}, this statistic can be viewed as a conditional maximum likelihood estimator and arguments based on asymptotic theory supports its pertinence in this situation, see \cite{Hill}.

\subsection{Geodesic distances} 

A set of connected vertices with the corresponding edges, constitutes a component of the graph. The collection of components forms a partition of the graph. We identify the components of the network and compute their respective sizes. When the size of the largest component is much larger than the size of the second largest component, see section IV A in \cite{Newman03} and the references therein, one then refers to the notion of giant component. 

A geodesic path between two connected vertices $x$ and $y$ is a path with shortest length that connects them, its length $d(x,y)$ being the geodesic distance between $x$ and $y$. One also defines the mean geodesic distance:
\begin{equation*}\label{eq:dist}
\mathcal{L}=\frac{1}{n(n+1)}\sum_{(x,y)\in \mathcal{V}^2}d(x,y),
\end{equation*}
where $\mathcal{V}$ denotes the set of all vertices of the connected graph and $n$ its size. For non-connected graphs, one usually computes a harmonic average. 
Mean geodesic distances measure ``how far" two randomly chosen vertices are, given the network structure. When $\mathcal{L}$ is much smaller than $n$, one says that a ``small-world effect" is observed. In this regard, the diameter of a connected graph, that is to say the length of the longest geodesic path, is also a quantity of major interest:
\begin{equation*} \label{eq:delta}
\delta=\max_{(x,y)\in \mathcal{V}^2}d(x,y).
\end{equation*}
Computations have been made for each component of the network of sexual contacts among individuals diagnosed as HIV positive before 2006 in Cuba, using the dedicated ``burning algorithm" for the mean geodesic distances, see \cite{AMO93}.



Along these lines, we also investigate how the connectivity properties of the network evolve when removing various fractions of specific strata of the population and studied the resilience and articulation points of the graph in Section 6 of the ESM.



\subsection{Mixing patterns} 

 We compute assortative mixing coefficients in order to highlight the possible existence of selective linking in the network structure. Various measures have been proposed in the literature for quantifying the tendency for individuals to have connections with other individuals that are similar in regards to certain attributes, depending on the nature of the latter (quantitative vs. qualitative). Precisely, given a (discrete) covariate, the population can be stratified into a partition $\mathcal{P}$ of $J\geq 1$ groups indexed by $j\in \{1,\; \ldots,\; J\}$ and one may calculate the proportion $m_{i,j}$ of edges in the graph connecting a node lying in group $i$ to another one in group $j$, $1\leq i \leq j\leq J$ and build the $J\times J$ mixing matrix $\mathcal{M}=(m_{i,j})$ (notice it is symmetric since edges are not directed here).
 Assortative mixing, as measured by $\mathcal{M}$, is then summarised by the modularity coefficient $Q_{\mathcal{P}}$ (e.g. \cite{newman_girvan_PRE2004}):
\begin{equation}
 Q_{\mathcal{P}}=Tr(\mathcal{M})-\vert\vert\mathcal{M}^2\vert\vert=\sum_{i}\left\{m_{i,i}-\left(\sum_{j=1}^N m_{i,j}\right)^2\right\}, \label{eq:modularity}
 \end{equation}where $\vert\vert A\vert\vert=\sum_i\sum_j a_{i,j}$ denotes the sum of all the entries of a matrix $A=(a_{i,j})$ and $Tr(A)$ its trace when the latter is square. We also define the assortative coefficient $r=Q_\mathcal{P}/(1-\vert\vert \mathcal{M}^2\vert\vert)$. As pointed out in \cite{Newman_assort}, large values of $r$ indicate "selective linking": it ranges from $0$ (randomly mixed network) to $1$ (perfectly assortative network).


Pursuing the analysis of possible ``preferential attachment", we investigate assortative mixing by vertex degree. In other terms, we tackle the question of whether the highly connected individuals have sexual contacts with other highly connected individuals. This amounts to study the linear correlation coefficient between degrees of adjacent vertices. We also plot the joint distribution of the pair indicating the respective numbers of contacts of two connected vertices. Additionally, the hypothesis of independence is tested by means of a standard $\chi^2$ method.


Other statistics are computed to illustrate the connectivity of the network: extremal dependence coefficients between the degrees of connected individuals and clustering coefficients to measure the transitivity of the graph (see ESM). Notice that because heterosexual men and MSM do not have sexual contacts, we expect naturally the graph to have a low clustering coefficient.



  Finally, we count and identify the maximal cliques of the network, i.e. subset of vertices, all adjacent, and such that there is no clique into which it is strictly included.

\subsection{Visual-mining}\label{subsec:visual-mining} 
Graph visualization techniques are used routinely to gain insights about
medium size graph structures, but their practical relevance is questionable
when the number of vertices and the density of the graph are high both for
computational issues (as many graph drawing algorithms have high complexities)
and for readability issues
\cite{DiBattistaEtAl1999GraphDrawing,HermanEtAl2000Graph}. Here, the situation
is borderline as the giant component of the graph contains 2,386 vertices and
3,168 edges (respectively 44.28\% and 77.78\% of the global
quantities). As the graph is of medium size from a computational point of view
and has a low density, it is a reasonable candidate for state-of-the-art
global and detailed visualization techniques. We use the optimised force
directed placement algorithm proposed in \cite{tunkelang_PHD1999}. It
recasts the classical force directed paradigm
\cite{fruchterman_reingold_SPE1991} into a nonlinear optimization problem in
which the following energy is minimised over the vertex positions in the
euclidean plane, $(z_1,\ldots,z_n)$,
\begin{equation*}
\mathcal{E}(z_1,\ldots,z_n)=\sum_{1\leq i\neq j\leq n}\left(a_{i,j}\frac{1}{3\delta}\|z_i-z_j\|^3-\delta^2\ln\|z_i-z_j\|\right),
\end{equation*}
where, $\delta$ is a free parameter that is roughly proportional to the
expected average distance between vertices in the plane at the end of the
optimization process, $a_{i,j}$ are the terms of the adjacency matrix of the network
 and $\|.\|$ denotes the Euclidean distance in the plane.

However, the structure of the graph under study, in particular its uneven
density, has adverse effects on the readability of its global
representation. We rely therefore on the classical simplification approach
\cite{HermanEtAl2000Graph} that consists in building a clustering of the
vertices of the graph and in representing the simpler graph of the
clusters. More precisely, the general idea is to define a partition composed of groups with dense internal links but low inter-group connectivity. Each group can then be considered as a vertex of a new graph: two
such vertices are connected if there is at least one pair of original vertices
in each group that are connected in the original graph.

Following
\cite{clemencon_etal_IWANN2011,clemencon_etal_ESANN2011,rossivilla-vialaneix2011societe-fran-caise},
we compute a maximal modularity clustering \cite{newman_girvan_PRE2004} as the
obtained clusters are well adapted to subsequent visual representation, as
shown in \cite{Noack2009}. An optimal $J$ classes partition,
$\mathcal{P}=C_1,\;\ldots,\; C_J$, is obtained by maximizing the modularity measure \eqref{eq:modularity}.
This optimization problem is NP-Hard and can only be solved via some
heuristics. As in \cite{rossivilla-vialaneix2011societe-fran-caise}, we use a
modified version of the multi-level greedy merging approach proposed in
\cite{NoackRotta2009MultiLevelModularity}: our modification guarantees that
the final clusters are connected. The optimization
process is carried out on the partitions for a given number of clusters $J$ but
also over the number of clusters $J$ itself which is then automatically selected. This makes the method essentially parameter free.

It should be noted however that one can find partitions with a rather high
modularity even in completely random graphs where no modular structure
actually exists (see \cite{ReichardtBornholdt2007ModularityDistribution} for
an estimation of the expected value of this spurious modularity in the limit
of large and dense graphs). To check that the modular structure found in a
network cannot be explained by this phenomenon, we use the simulation
approach proposed in
\cite{clemencon_etal_IWANN2011,clemencon_etal_ESANN2011,rossivilla-vialaneix2011societe-fran-caise}. Using a Monte Carlo Markov Chain (MCMC) approach inspired by \cite{Roberts2000MCMC},
we generate configuration model graphs with exactly the same size and degree distribution as the epidemics graph. Using the above algorithm, we compute a
maximal modularity clustering on each of those graphs. The modularities of the
clustering provide an estimate of the distribution of the maximal
modularity in random graphs with our degree distribution. If a partition of this graph
exhibits a higher modularity, we conclude that it must be the result of
some actual modular structure rather than a random outcome.

The maximal modularity clustering is visualised using the force
directed placement algorithm described above. In addition to giving a general
idea of the global structure of the graph, the obtained visual representation
can be used to display distributions of covariates at the cluster level. Homogeneity tests are performed in order to
assess possible significant differences between these statistical
subpopulations.

However, as demonstrated in \cite{FortunatoBarthelemy2007}, finding the
maximal modularity clustering can lead to ignoring small modular structures
that fall below the resolution limit of the modularity measure. It is then
recommended in \cite{FortunatoBarthelemy2007} to recursively apply maximal
modularity clustering to the original clusters in order to investigate
potential smaller scale modules. We follow this strategy coupled with the
MCMC approach described above: each cluster is tested for substructure by
applying the maximal modularity clustering technique from
\cite{rossivilla-vialaneix2011societe-fran-caise} and by assessing the actual
significance of a potential sub-structure via comparison with similar random
graphs.

\section{Results and Discussion}\label{sec:results}

\subsection{Overview of the Cuban HIV epidemics with contact-tracing}\label{subsec:overview} 

We start off with a brief review of some macroscopic statistics (on the basis of Sections 2 and 3 of the ESM, to which we refer for further figures and details).

\begin{figure}[H]
\begin{center}
\includegraphics[width=10cm,height=8.5cm]{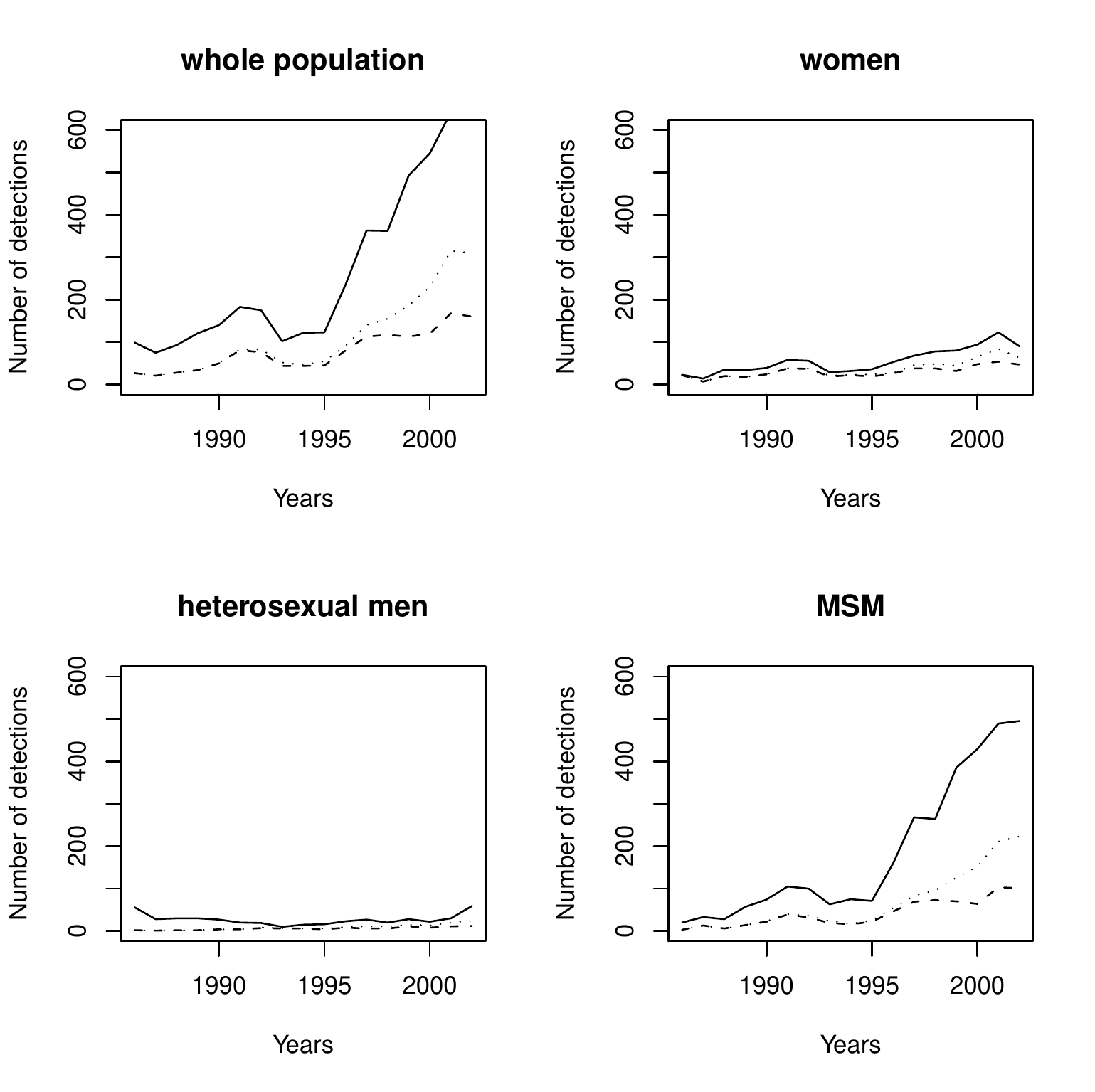}\\
     \caption{{\small Yearly sexual distribution of newly detected individuals. For each plot, the plain line corresponds to the total number of new detections, the dashed line to the individuals detected by contact-tracing and the dotted line to the individuals detected by non-random methods (contact-tracing or captation).}}       \label{fig:evolution}
\end{center}
\vspace{-0.5cm}
\end{figure}

Most HIV infections in Cuba occurred among MSM individuals (68.92\%) and among people living in Ciudad de la Habana (53.48\%). However, the evolution of the number of HIV detections within groups of different sexual orientations, Fig. \ref{fig:evolution}, reveals that, at the very beginning of the epidemic, infections occurred predominantly in the heterosexual men subpopulation. In the following years, women and then MSMs were then contaminated. One also observes a significant increase of the reported HIV diagnoses, starting from 1995. Further analysis is carried out in Section 4 of the ESM by studying the infection tree, and a film showing the emergence of the giant component is also given in the supplementary materials. Age distributions show a slight ageing of newly infected individuals. Over the observation period 1986-2006, the mean age at detection is 29.43 years for the whole population, while it is lower for women.\\

There are several ways of HIV detection: random screening, contact-tracing or captation (i.e. following a medical examination by the family doctor). The proportion of new HIV cases diagnosed thanks to contact-tracing detection continually increased over the observation period. Among the individuals detected as HIV positive between 1986 and 2006, 1,515 (respectively 821 and 3,053) persons have been detected by contact-tracing (respectively by captation and random-screening methods), which amounts to 28\% (respectively 15\% and 57\%) of the total number of HIV detections. The ``nonrandom detection methods" (i.e. detections by contact-tracing or captation) account for 65.5\%, 33.5\% and 38.2\% of detections for women, heterosexual men and MSM respectively.
A $\chi^2$-test reveals that the variables 'gender/sexual orientation' and 'way of detection' are not independent, even when restricted to a geographical region. The strong departure from independence is due to the following facts: 1) More women are detected by contact-tracing than by random detection. 2) More men are detected through random detections than by contact-tracing. 3) Detection by captation is as efficient whatever the sexual orientation.\\
Contact-tracing is most efficient for women and for individuals living in the West and Central areas. In addition, women contribute most to the reporting of new diagnosed cases: they are responsible for 29.66\% of the positive diagnoses following contact-tracing.\\

In \cite{arazozaclemencontran} and \cite{blumtran}, statistical analysis at the population level shows that, in the last years, 'contact-tracing' and 'captation' methods accounted for almost one detection out of two. Here, we complete this analysis by exploiting the individual and network data.
When an individual gives her/his contacts, on average 77\% of the latter are tested (median of 85\%) and 22\% of the contacts given by an individual are tested positive (median of 14\%). It seems that investigating the contacts of an individual detected by contact-tracing is an efficient way of finding new infected individuals. The proportion of tested (respectively diagnosed HIV positive) contacts for individuals who have been detected by contact-tracing is 80.91\% (resp. 34.15\%), which is higher than the proportion for the whole population (76.89\% resp. 22.52\%) or for individuals detected by random methods (76.29\% resp. 18.57\%).


To have an idea of how the non-random detections help the health system to detect new infectious individuals, time of detection along one edge is investigated. Detection times when the individual (ego) has been detected by contact-tracing or captation, are shown to be shorter (699 and 428 days) compared with the times for egos detected by random methods (726 days). As these differences could be explained by the fact that individuals detected by non-random methods belong to clusters that are actively explored, we also consider only the edges that have served for contact-tracing detections and obtain the same conclusions.



\subsection{Analysis of the degree distribution.}\label{subsec:degree} 


\begin{figure}[H]
\begin{center}
\includegraphics[width=10cm,height=4.5cm]{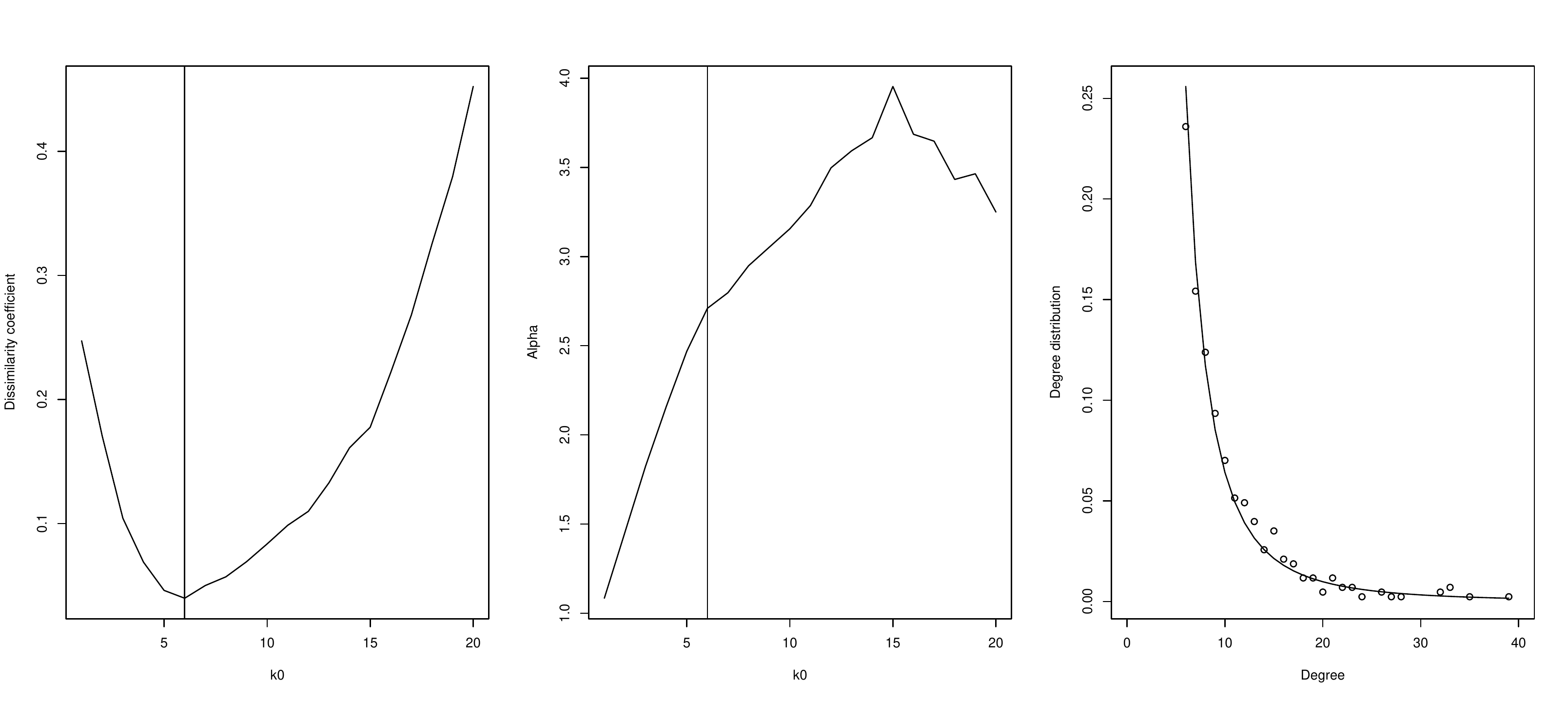}\\
     \caption{{\small Observed degree distribution for the whole HIV+ population and estimated power law.}}       \label{fig:heavy}
\end{center}
\vspace{-0.5cm}
\end{figure}





Among the 5,389 individuals appearing in the database, 483 declared no sexual partners during this period. Degree distributions for the whole population exhibit a clear power-law behavior. Power laws are fitted to the declared degree distributions, for the whole population and for the strata defined by the variable gender/sexual orientation respectively. Both methods present similar results. 
The resulting estimates reveal the thickness of the upper tails: we obtain $(k_0,\; \alpha)=(7,\; 3.06)$ for the pooled population, $(7,\; 3.02)$ for MSM, $(6,\; 2.71)$ for women and $(7,\; 3.36)$ for heterosexual men. Women correspond to the heaviest tail, followed by MSM and heterosexual men (the smaller the tail exponent $\alpha$, the heavier the distribution tail). However, an ANOVA reveals no statistically significant impact of the covariates gender/sexual orientation, detection mode and area of residence on the declared degree distribution, although differences can be seen on the estimates. When considering the influence of the variable age at detection on the degree, we find a significative slope of $-0.1$ but with a very small determination coefficient.  This is explored further in the ESM.

All the same, using the observed degree distribution, we obtain $(k_0,\; \alpha)=(3,\; 2.99)$ which is very close to the result when using the number of neighbours having been detected positive. \\
All the tail exponent estimates are below the critical value of $\alpha_c=3.4788$, below which a giant component exists in scale-free networks generated by means of the configuration model, and above the value $2$, below which the whole graph reduces to the giant component with probability one.

\paragraph{Joint degree distribution of sexual partners.}

\begin{figure}[H]
\begin{center}
      \includegraphics[width=6cm]{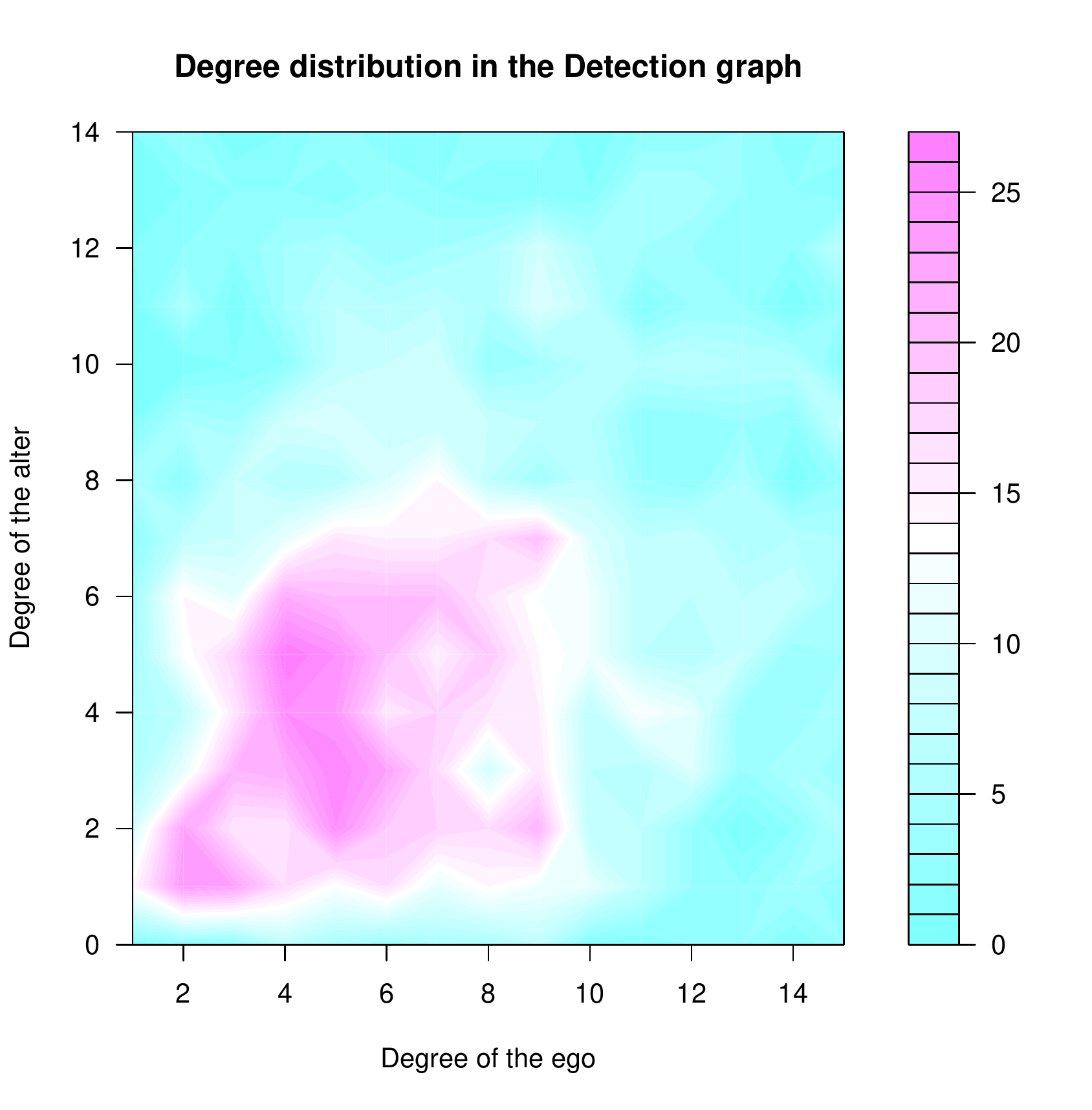}\\
     \caption{{\small\textit{Joint degree distribution of the number of contacts for connected vertices.}}}
     \label{fig:joint_distrib_degree}
\end{center}
\vspace{-0.5cm}
\end{figure}

The independence assumption between the degrees of adjacent vertices does not hold here, see Fig. \ref{fig:joint_distrib_degree}, in contrast to what is assumed for the vast majority of graph-based SIR models of epidemic disease, e.g. \cite{durrett, Newman03}. Indeed, the linear correlation coefficient between the degree distribution of alters and egos is equal to $0.68$ and independence is rejected by a $\chi^2$-test with a p-value of $6.85\, 10^{-6}$. In particular, highly connected vertices tend to be connected to vertices with a high number of connections too. 
From the perspective of mathematical modeling, this suggests to consider graph models with a dependence structure between the degrees of adjacent nodes, in opposition to most percolation processes on (configuration model) networks used to describe the spread of epidemics \cite{MolloyReedCriticalPoint1995, ballneal, volz, decreusefonddhersinmoyaltran, grahamhouse}. However, it is worth noticing that, if we restrict our analysis to some specific, more homogeneous, subgroups, the independence assumption may be grounded in evidence. So if assumptions such that the network is generated by a configuration model do not hold globally, they may be valid for smaller clusters, which is another motivation for clustering.



\paragraph{Preferential attachment.} 

The network clearly exhibits a phenomenon of ``assortative mixing" with respect to the area of residence, age, detection mode: individuals sharing these characteristics are connected or have short connecting paths. The vast majority of connections unsurprisingly occur between individuals in the same geographical region, and among the latter, 1,678 ($41$\%) link two individuals living both in the Ciudad de la Habana, resulting in a very high assortative mixing coefficient, 0.7753. Those who are between 20 and 40 at the time of detection make 61.16\% of the edges of the whole graph and 62.88\% of the edges in the infection tree. Among these edges, 28.32\% (resp. 25.80\% in the infection tree) link individuals of the same group $[20,25)$, $[25,30)$, $[30,35)$ or $[35,40)$. Other relations and average geodesic distances in the giant component with respect to the various attributes are computed in the ESM.

As edges correspond to sexual contacts in the present graph, the gender/sexual orientation of adjacent vertices cannot be arbitrary of course. When considering the sexual orientation of two neighbours, we see that more than a half of the edges (56.47\%) link two MSM. Links between MSM and women make 1,208 edges (29.66\%) and there are 439 edges (10.78\%) between women and heterosexual men. Looking at the infection tree provided similar proportions: 1,202 edges (52.56\%), 667 edges (29.16\%) and 375 edges (16.40\%) respectively. Figures reveal an asymmetry in HIV infection: among (oriented) infection edges involving women, the latter are more often alters than egos (66.13\% of the edges shared with heterosexual men and 74.21\% of the edges shared with MSM).
The declarative degree shows a smaller mean degree for heterosexual men and comparable degree distributions between women and MSM. MSM are expected to contribute most to the connectivity of the graph, especially bisexual men who act as contact points between women and MSM who declare only contacts with men.

\subsection{Analysis of the "giant component"}\label{subsec:giant} 


Several connected components form the sexual contact network of the Cuban HIV database. One is much larger than the others with 2,386 HIV-infected individuals (44.28\%), while the other components count less than 17 vertices (see ESM). Incidentally, we underline that, although the number of non HIV-infected sexual partners of the individuals infected by the HIV between 1986 and 2006 is known, no information about possible sexual contacts between the latter (which could potentially lead to connect some of the disjoint components aforementioned) is available.\\
There are 1,627 isolated individuals, most of which are men, especially heterosexual men. The isolated couples are mostly composed of man-woman couples with an over-representation of women compared to the rest of the population. The main reasons why these individuals are isolated are that they have few positive contacts and there is a high fraction (22.06\%) who did not provide their contacts (8.96\% and 2.80\% in the whole population and in the giant graph resp.).


The network density is globally low and very heterogeneous. But although the connectivity of the network seems fragile at first glance, density may be locally very high. The harmonic average of the geodesic path lengths equals 10.24 and 12.2 for the oriented graph. Most of the graph connectivity is concentrated in the largest component (3,168 edges out of 4,073). The largest component has a diameter of 26 (36 when taking into account the direction of the infections) and the harmonic average of the geodesic path lengths are the same inside the largest component. These values are slightly higher than those of other real networks mentioned in \cite{Newman03} but remain well below the number of vertices and compatible with the logarithmic scaling related to the so-termed small world effect.

\begin{figure}[htbp]
  \centering
\includegraphics[width=0.9\linewidth]{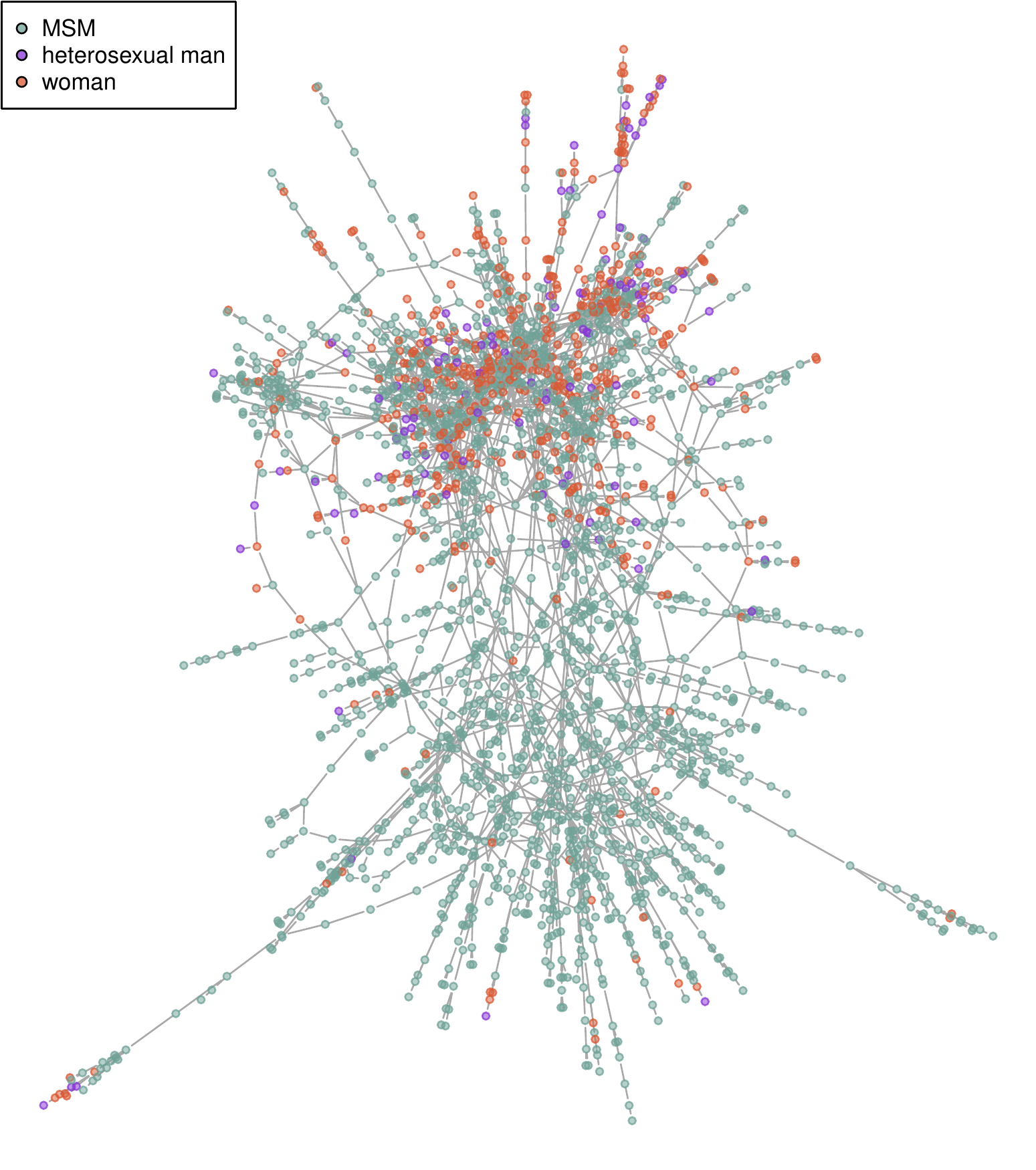}
  \caption{Force directed placement layout of the largest connected component}
  \label{fig:cuba_largest_cc_fdp}
\end{figure}

Figure \ref{fig:cuba_largest_cc_fdp} shows the layout obtained by means of the
force directed placement algorithm from
\cite{fruchterman_reingold_SPE1991,tunkelang_PHD1999}. The general picture
seems quite clear, with what appears to be two parts in the graph: the
lower part of the graph (on the figure) seems to be dominated by MSM while
the upper part gathers almost all persons from the giant component that have
only heterosexual contacts. However, the upper part is quite difficult to read
as it seems denser than the lower part. The
layout shows what might be interpreted as cycles and also a lot of small trees
connected to denser parts. The actual connection patterns between the upper
part and the lower part are also very unclear. Because of these crowding
effects, structural properties of the network from Figure
\ref{fig:cuba_largest_cc_fdp} appears quite difficult and probably
misleading. We rely therefore on the simplification technique outlined in
Section \ref{subsec:visual-mining} leveraging a clustering of the giant
component to get an insight into its general organization. 

\paragraph{A clustering of high modularity.} 

\begin{figure}
\begin{center}
\includegraphics[width=0.9\linewidth]{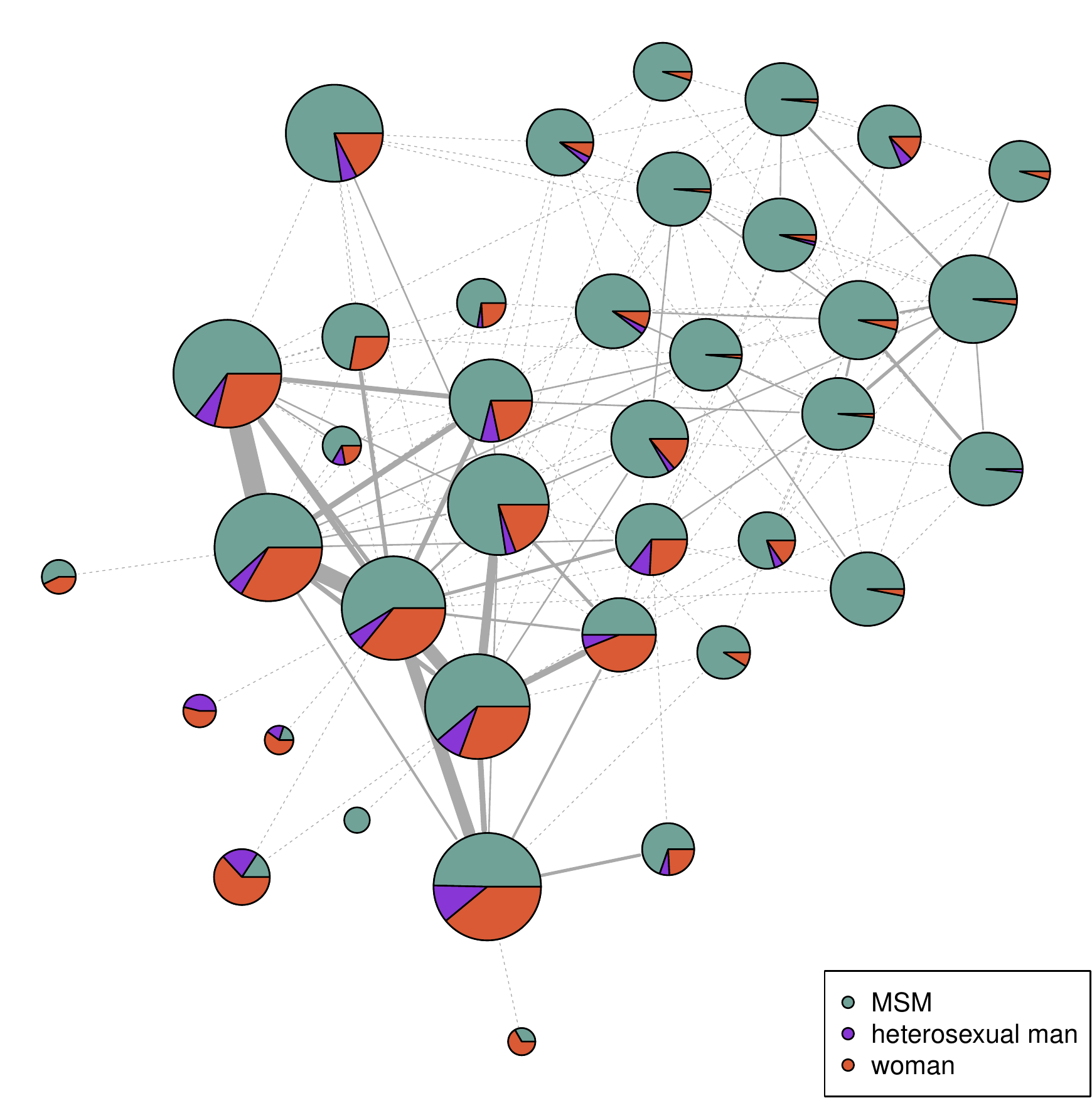}
     \caption{{\small\textit{The giant component divided into 37 clusters. Each disk of
       representation corresponds to one cluster and has an area proportional
       to the number of persons (original vertices) gathered in the associated
       cluster. The pie chart of the disk displays the percentage of MSM,
       of heterosexual men and of women in the cluster. Links between clusters
       summarise the connectivity pattern between members of the clusters. The
       thinnest edge width corresponds to only one connection between a member
       of one cluster and another person in the connected cluster (the
       corresponding edges are drawn using dashed segments). Thicker
       edges have a width proportional to the number of connected persons.}}
     }
     \label{fig:cluster1}
\end{center}
\end{figure}

A graphical representation of the partition obtained by the
method from \cite{rossivilla-vialaneix2011societe-fran-caise} is displayed in
Figure \ref{fig:cluster1}. The clustering thus produced exhibits a
modularity of 0.8522 and is made up of 37 clusters.
This modularity is
very high compared to the random level and strongly supports the hypothesis of
a specific (``non random") underlying community structure. For comparison
purpose, the average maximal modularity attained by random graphs built from a
configuration model with the same size and degree distribution as those of the
giant component observed over a collection of 100 simulated replications
(using the same partitioning method) is of the order 0.74, with a maximum of
0.7435.

Considering that the modules are meaningful, the visual representation
provided by Figure \ref{fig:cluster1} is more faithful to the underlying
graphical structure than the finer displays of Figure
\ref{fig:cuba_largest_cc_fdp}. That said, the two graphs tend to agree as the pie
charts of Figure \ref{fig:cluster1} clearly show two parts in the network: the
lower left part seems to gather most of the women and heterosexual
men (as the upper part of Figure \ref{fig:cuba_largest_cc_fdp}), while the
upper right part contains clusters made almost entirely of MSM, as the lower
part of Figure \ref{fig:cuba_largest_cc_fdp}. While the display of Figure
\ref{fig:cluster1} might seem cluttered, it is in fact very readable if one
considers that only 328 edges of the giant component connect persons from
different clusters while 2,840 connections happen inside clusters. Then most of
the edges on Figure \ref{fig:cluster1} could be disregarded as they
corresponds to only one pair of connected persons (this is the case of 94 of
such edges out of 142 and the former are represented as dashed segments). Taken
this aspect into account, it appears that the MSM part of the giant component
(upper right part) is made of loosely connected clusters while the bulk of the
connectivity between clusters is gathered in the mixed part of the component,
in which most women and heterosexual men are gathered. The fact that the mixed part is more dense was already visible in Figure
\ref{fig:cuba_largest_cc_fdp}, but Figure \ref{fig:cluster1} provides a much
stronger demonstration.

\paragraph{Inhomogeneous covariate distributions.} 
The pie chart based visualization of Figure \ref{fig:cluster1} shows the sexual orientation distribution in the clusters and hence sheds light on its relationship with the graphical
structure. To get a more precise assessment of those links, we especially analyse the
homogeneity of the clusters with respect to the covariate
`gender/sexual orientation'.

As pointed out before, the general layout of the largest connected component
displayed in Figure \ref{fig:cluster1} suggests that the global structure can
be viewed as the union of two distinct parts: one part contains women and
heterosexual men in majority (lower left region of the Figure), while the
other part is mostly composed of MSM (upper right region). This fact can be
strongly supported by examining the distribution of the variable
gender/sexual orientation depending on the cluster; see Section 7 in the ESM, where the clusters are
sorted by increasing order of magnitude of the p-value of a $\chi^2$ test of
homogeneity, in order to test whether the sexual orientation distribution in
the cluster coincides with that in the whole largest connected component.

\begin{figure}
\begin{center}
\includegraphics[width=0.9\linewidth]{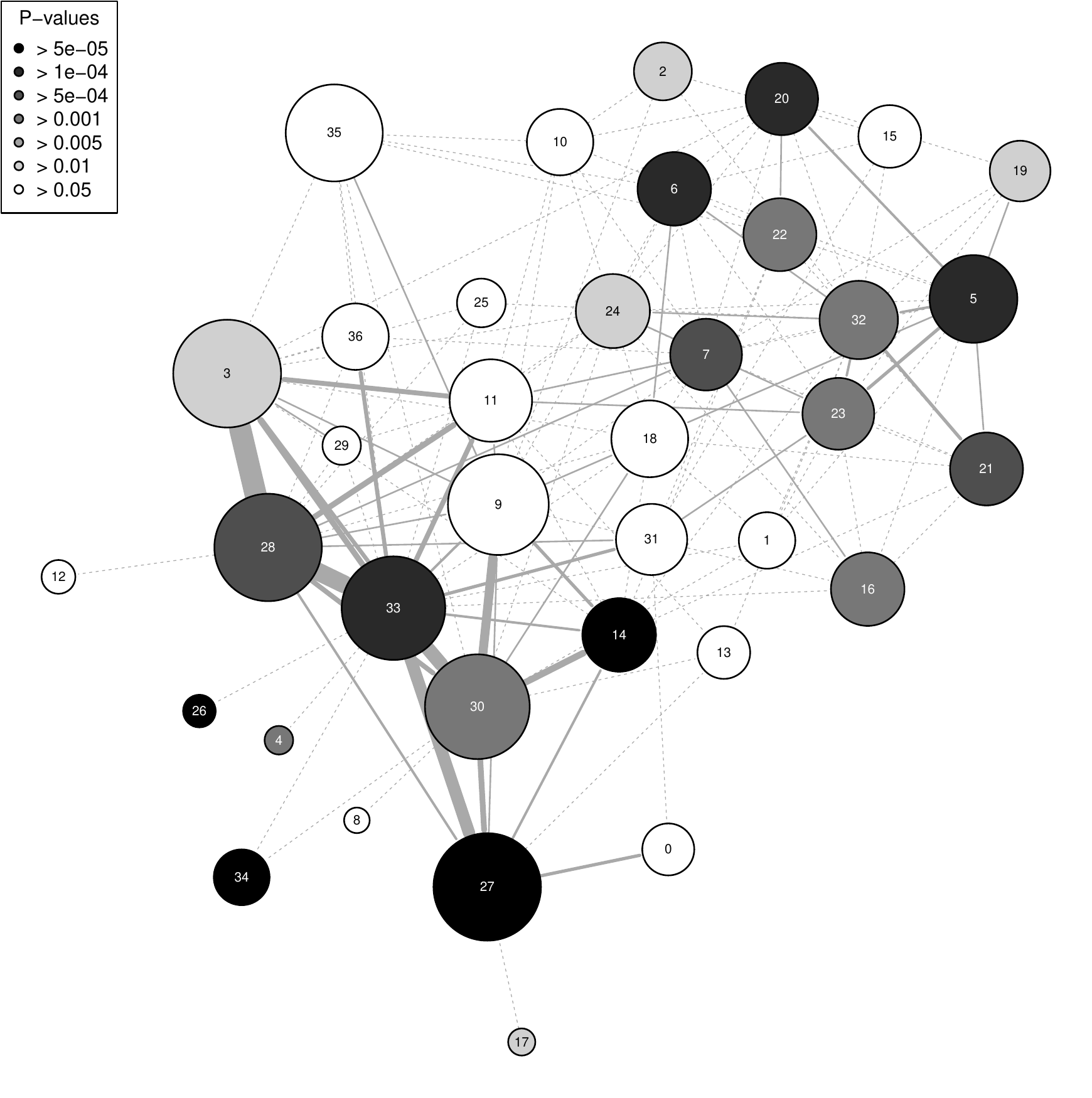}
     \caption{{\small\textit{The giant component divided into $37$ clusters. Disk areas and
       edges thicknesses are chosen as in Figure \ref{fig:cluster1}. The grey
       level of a disk encode the $p$-value of a $\chi^2$ test of
       homogeneity in which the distribution of the sexual orientations in the
     associated cluster is compared to the distribution in the giant component.}}}
     \label{fig:SO:clustered:chisq}
\end{center}
\end{figure}

Figure \ref{fig:SO:clustered:chisq} gives a visual representation
of the corresponding $p$-values: the darker the node, the more statistically significant
the difference between the cluster distribution of sexual orientation and the
distribution of the giant component. 
For instance, cluster 35 (a large white vertex in
the upper left part of Figure \ref{fig:SO:clustered:chisq}) has a typical
sexual orientation distribution. 
Combining Figures
\ref{fig:cluster1} and \ref{fig:SO:clustered:chisq} is very useful: Figure
\ref{fig:SO:clustered:chisq} highlights atypical clusters while Figure
\ref{fig:cluster1} identifies why they are atypical. For instance, cluster 27
is atypical because it contains both a lot of women and of heterosexual men,
while cluster 6 is atypical because it contains almost only MSM.

\begin{figure}
\begin{center}
\includegraphics[width=0.9\linewidth]{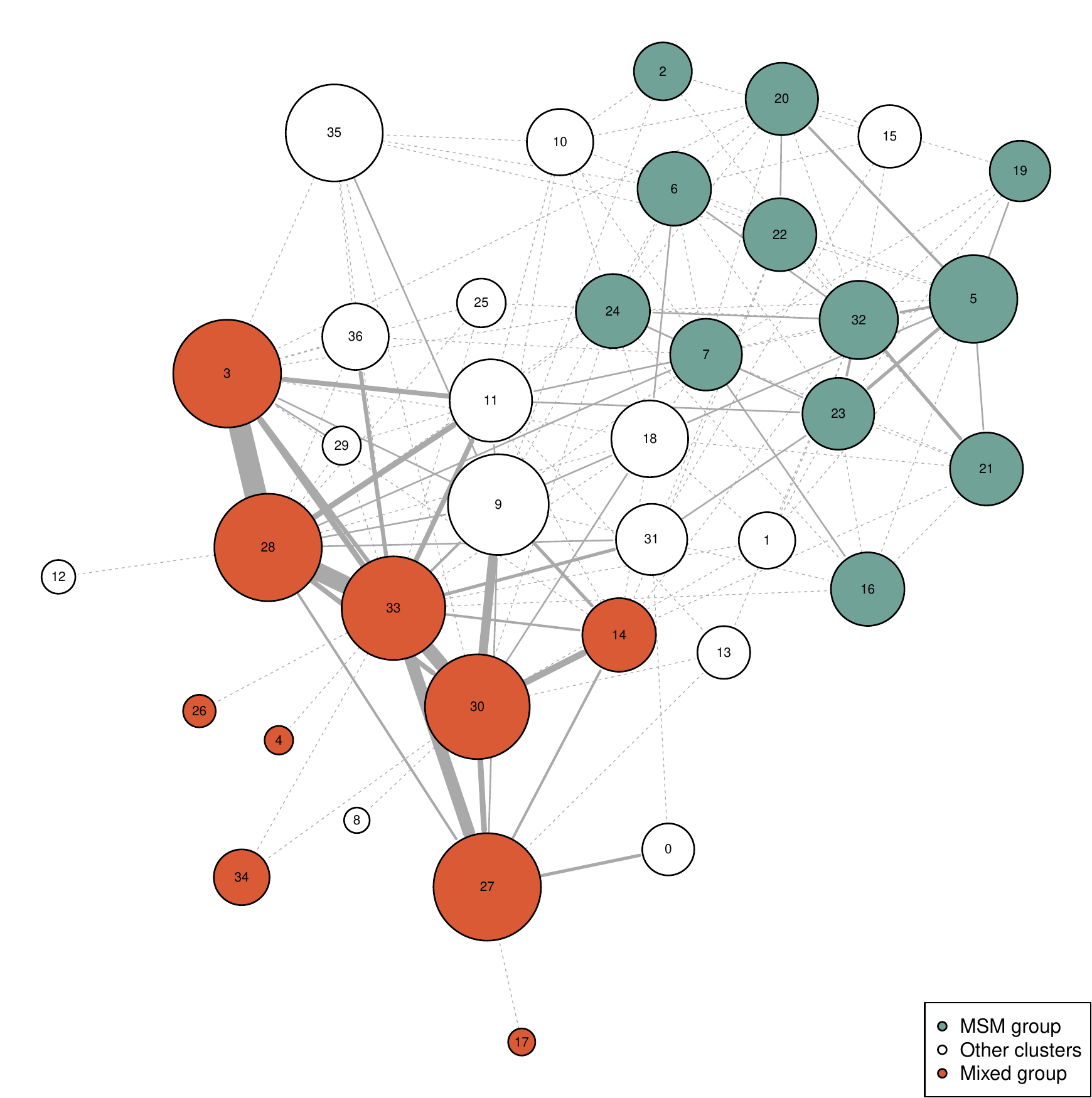}
     \caption{{\small\textit{Three groups of clusters in the giant component. White clusters
       are typical, while green ones are dominated by MSM and red ones
       contain a larger number of heterosexual persons than expected according
     to the global distribution of sexual orientations in the giant component.}}}
     \label{fig:SO:clustered:groups}
\end{center}
\end{figure}

It appears that among the 37 clusters, 22 exhibit a $\chi^2$ p-value below
5\%. They will be abusively referred to as ``atypical clusters" in the
following. The set of those clusters can be split into two subsets, depending
on the percentage of MSM in the cluster: above or below the global value of
76\% (the percentage in the giant component), as illustrated by Figure
\ref{fig:SO:clustered:groups}. Almost two thirds (67\%) of the
individuals of the largest connected component lie in the atypical
clusters. Among the latter, 774 individuals belong to the 12 clusters which
display a large domination of MSM (denoted the MSM group of clusters in the
sequel) and 825 to the 10 clusters that contain an unexpectedly large number
of heterosexual persons (denoted the mixed group of clusters in the
sequel). See Section 7 of the ESM for detailed population numbers.

According to Figure \ref{fig:SO:clustered:groups}, the two subsets of atypical
clusters seem to be almost disconnected. This is confirmed by a detailed
connectivity analysis. There are indeed 864 internal connections in the MSM group,
1,276 in the heterosexual group, and only 10 links between pairs of
individuals belonging to the two different groups. This asymmetry was expected, given the quality of the clustering with only 328 inter-cluster connections. Nevertheless, the number of
connections between the two groups of clusters is also small compared to
connections between the clusters of the groups: 129
connections between persons of distinct clusters in the group of mixed
clusters and 55 in the group of MSM clusters. Finally, there are 83
connections from persons in the group of mixed clusters to persons in non
atypical clusters, and 36 connections from persons in the group of MSM
clusters to persons in non atypical clusters. Mean geodesic distances inside the MSM group are larger than in the mixed
group (respectively 9.95 and 7.28, computed without orientation). To conclude, the two groups are weakly connected to the outside, with a small number of direct connections, and rather internally more connected than expected.



\paragraph{Sub-structures.} 

\begin{figure}[htbp]
  \centering
  \begin{tabular}{cc}
\includegraphics[scale=0.4]{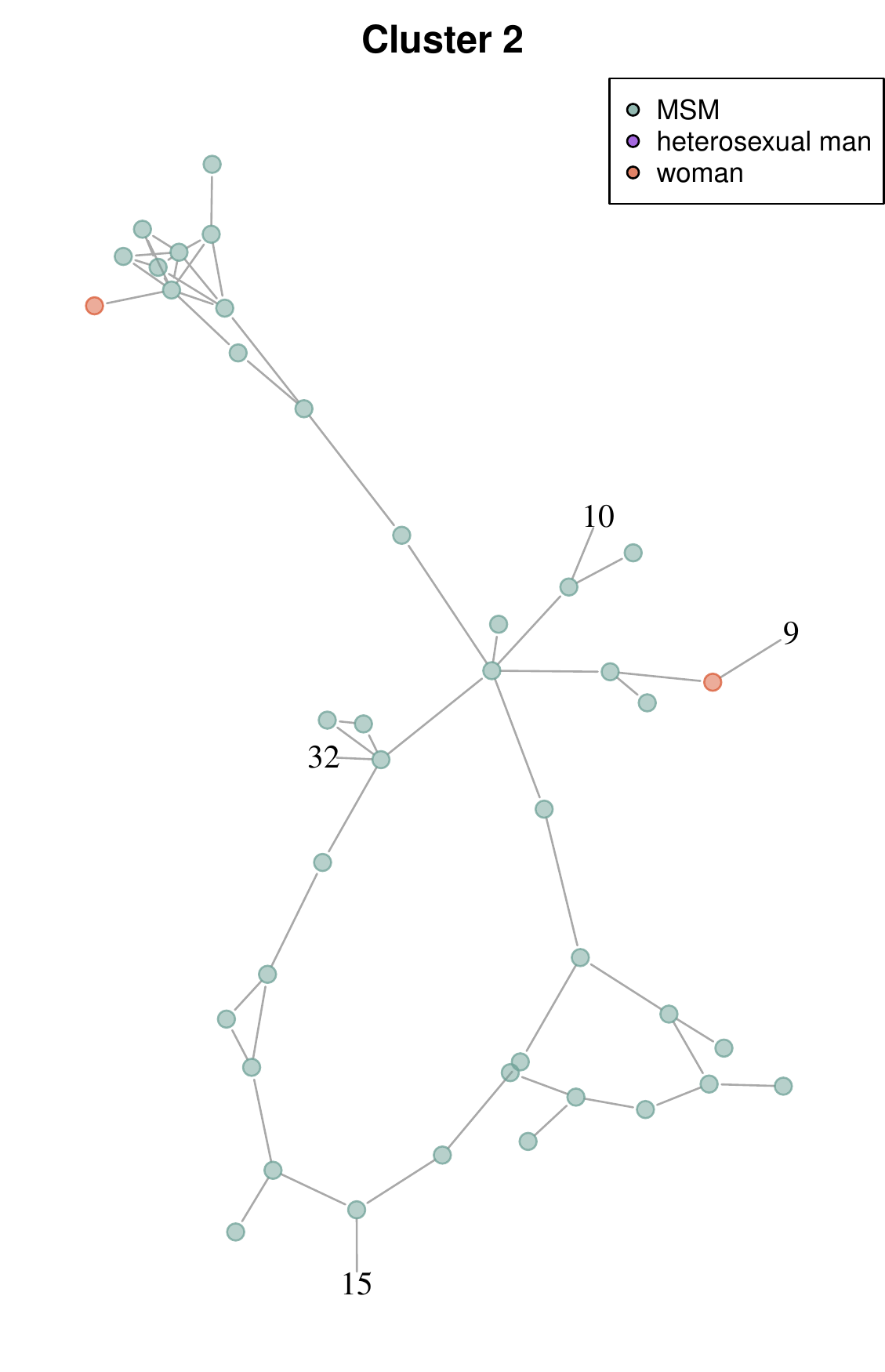}&
\includegraphics[scale=0.4]{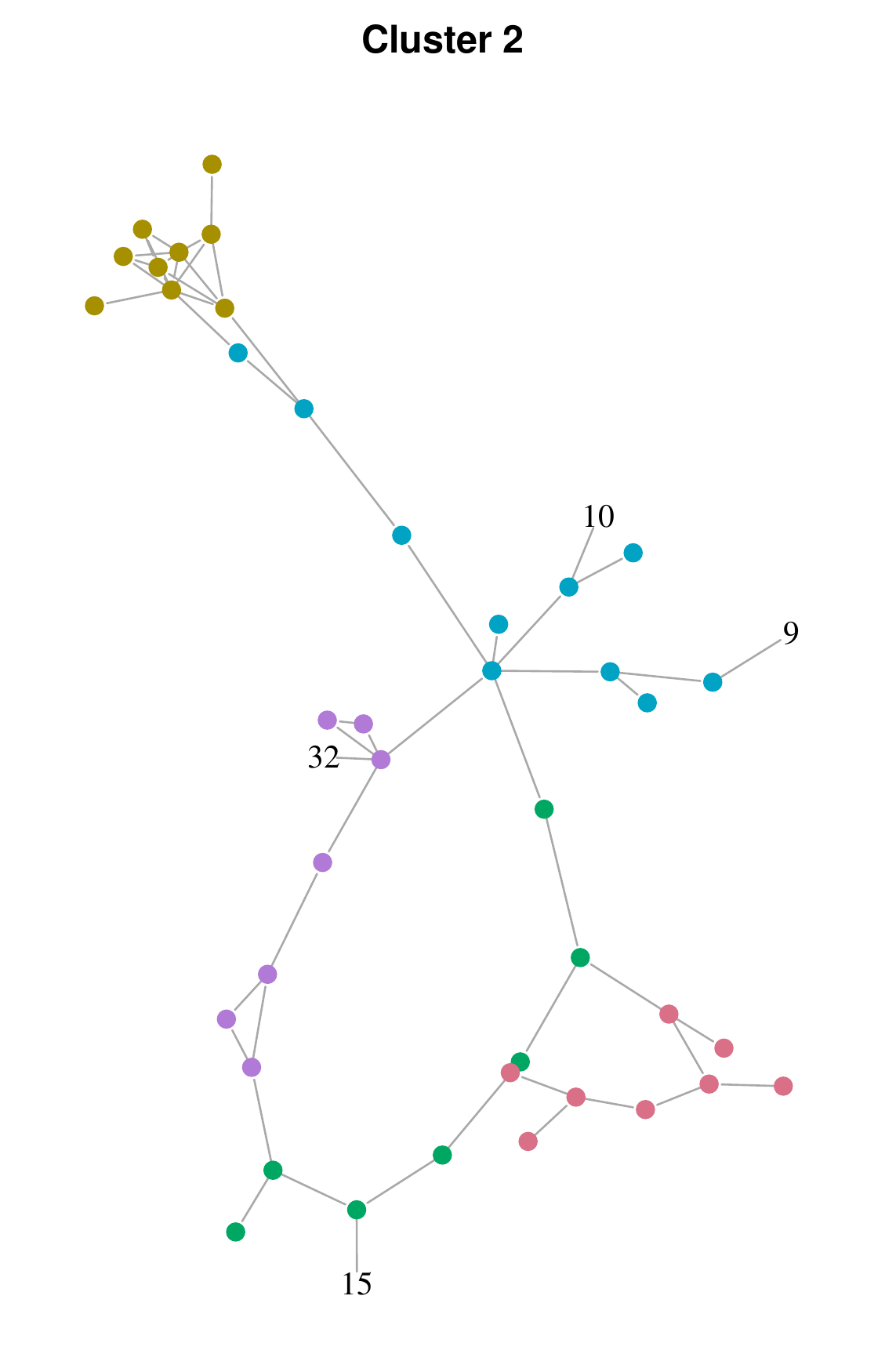}\\
  \end{tabular}
  \caption{{\small\textit{A visual representation of Cluster 2. The left sub-figure shows the sexual orientation of the persons while the right one shows the five sub-clusters (color coded) as discovered by the recursive maximal modularity clustering. Numbered nodes represent clusters to which persons in Cluster 2 are connected.}}}
  \label{fig:Cluter2:sub}
\end{figure}

According to \cite{FortunatoBarthelemy2007}, the
resolution limit of the modularity is given by $\sqrt{2m}$ where $m$ is the
total number of edges of the graph under analysis. In the largest connected
component, the resolution limit is roughly 80: as shown in the ESM, 28 out of 37 clusters have less than 80
edges and should therefore be investigated. Using the procedure described in Section
\ref{subsec:visual-mining}, we find that only 13 clusters have a
sub-structure, and that among them, 6 are above the resolution limit. For the 7 ones below the resolution limit, the sub-structures
are not very enlightening as shown on Figure \ref{fig:Cluter2:sub} for a typical example.

The partition in Figure \ref{fig:Cluter2:sub} shows a typical
behavior of maximal modularity clustering: while the upper left cluster shows
a densely connected subgraph of MSM (with a woman), the other clusters are
infection chains separated by vertices with a degree slightly above the
average. For instance, the blue central cluster can be separated from the
others because the connection vertex has a degree six and is connected with a
purple vertex with degree four (in the cluster). Then their connection is not
as significant as others in the cluster and can be cut. This shows the
practical limit of the clustering technique as a tool for guiding the analysis
of the contact graph. While displaying possible infection trees could be very
interesting, there is no reason for these tiny substructures to be those trees
as they are not constructed by taking detection dates or other information
into account. Moreover, in practice, most of the clusters are in general quite small and can be
easily visualised as shown in the ESM. Those figures show interesting patterns such
as quite long contact chains (see e.g. Cluster 10 in the ESM) and star like patterns around high degree nodes (see e.g.,
Cluster 23 and Cluster 32). Those patterns could analyse further manually in order
to understand their influence in the transmission of the disease, for instance.

In fact, only the largest clusters lead to complicated figures with
overlapping representation for which a sub-cluster analysis could be
useful. Those clusters are listed in Table \ref{tab:Complex:Clusters}.

\begin{table}[htbp]
  \centering
  \begin{tabular}{|cccccc|}\hline
Cluster & Vertices & Edges & Sub-clusters & p-value & type\\\hline
3 & 142 & 195 & 11 & 0.0122 & Mixed\\
5 & 94 & 101 & None & 0.0002& MSM\\
9 & 124 & 143 & None & 0.7597 & Typical \\
11 & 83 & 93 & 10 & 0.4439 & Typical \\
27 & 141 & 236 & 11 & 0.0001 &  Mixed \\
28 & 141 & 184 & 10 & 0.0007 &  Mixed \\
30 & 134 & 177 & 17 & 0.0014 & Mixed \\
33 & 131 & 214 & None & 0.0003 & Mixed \\
35 & 115 & 155 & 10 & 0.7890 & Typical\\
36 & 54 & 66 & None & 0.1130 & Typical\\\hline
  \end{tabular}
  \caption{{\small\textit{Clusters with complex graphical representations. The number of
    sub-clusters is given by a recursive application of the maximal modularity
  clustering. When the table gives ``None'', it means that the sub-clustering
  with consider non significant according to the random level of the
  modularity estimated with the MCMC approach. Results from the sexual
  orientation distribution analysis are recalled in the table for easier reference.}}}
  \label{tab:Complex:Clusters}
\end{table}

Interestingly, most of the complex clusters are from the Mixed and the Typical
groups of clusters. MSM clusters tend to be smaller or less complex: most of
their complexity has already been explained away by the top level
clustering. Additionally, four of those clusters have no sub-structure, which
means that they can be considered (at least from the modularity point of view)
as configuration model graphs (with some specific degree distribution), and that there is
no point in trying to find simple sub-structure in them. For these sub-clusters, predictions and approximations based on the assumption of a configuration model hold and can be used. 


\section{Conclusion}

The major contribution of the present article lies in this premier exploitation of individual data related to the sexual contacts from the Cuban HIV database. The latter gave us the opportunity to observe the sexual contact network through which HIV has spread in Cuba between 1986 and 2006. We computed informative empirical measurements related to its structure. Previous works devoted to the study of the spread of the HIV in Cuba ignored the underlying network structure, for instance see \cite{arazozaclemencontran,blumtran} for a stochastic modeling taking into account the effect of contact-tracing and \cite{Hector,hsiehwangarazozalounes} for a deterministic approach based on nonlinear partial differential equations. Here, a variety of interesting features of the HIV epidemic in Cuba have been highlighted by means of recent network mining methods, such as heavy-tailness of the degree distribution, occurrence of a giant connected component and statistically significant presence of a community structure. The network approach shows that investigating the contacts of contact-traced individuals is not less efficient nor longer than considering contacts of randomly detected ones. 
Whereas global statistics indicate a low density of the graph (many articulation points, resilient structure, low clustering coefficients), the clustering emphasised the  important heterogeneity in the network, with some dense regions that are internally more connected than average and with few links to the outside. We find subgroups with atypical covariate distributions, each reflecting a different stage of the evolution of the epidemic. Clustering the graph also allows us to unfold the complex structure of the Cuban HIV contact-tracing network. As a byproduct, the clustering indicates sub-structures that may be considered as random graphs resulting from configuration models, bridging the gap between the modelling papers whose assumptions on network structures do not often match reality. Finally, the present work is a valuable step towards building a mathematical model of HIV spread in presence of contact-tracing that generates graphs whose properties (partly) match those we listed in this paper.


\section*{Acknowledgements}
This work has been funded by ANR Viroscopy (ANR-08-SYSC-016-03), Chaire Math\'ematiques et Mod\'elisation de la Biodiversit\'e (Ec. Polytechnique, Museum National d'Histoire Naturelle et Fondation X), ANR MANEGE (ANR-09-BLAN-0215) and Labex CEMPI (ANR-11-LABX-0007-01). H. De Arazoza received support from the Spanish project AECID A2/038418/11. The authors thank Dr. J. Perez of the National Institute of Tropical Diseases in Cuba for granting access to the HIV/AIDS database. They also thank Ms. D. Abu Awad for reviewing the English language.


\section*{Short title for page headings}

Network analysis of the Cuban HIV epidemics
\end{document}